# Survey on Security Issues in Cloud Computing and Associated Mitigation Techniques

Rohit Bhadauria*
School of Electronics and Communications Engineering
Vellore Institute of Technology, Vellore, India
bhadauria.rohit@gmail.com

Sugata Sanyal
School of Technology and Computer Science
Tata Institute of Fundamental Research, Mumbai, India
sanyals@gmail.com

## ABSTRACT
Cloud Computing holds the potential to eliminate the requirements for setting up of high-cost computing infrastructure for IT-based solutions and services that the industry uses. It promises to provide a flexible IT architecture, accessible through internet from lightweight portable devices. This would allow multi-fold increase in the capacity and capabilities of the existing and new software. In a cloud computing environment, the entire data resides over a set of networked resources, enabling the data to be accessed through virtual machines. Since these data-centres may be located in any part of the world beyond the reach and control of users, there are multifarious security and privacy challenges that need to be understood and addressed. Also, one can never deny the possibility of a server breakdown that has been witnessed, rather quite often in the recent times. There are various issues that need to be addressed with respect to security and privacy in a cloud computing environment. This extensive survey paper aims to elaborate and analyze the numerous unresolved issues threatening the cloud computing adoption and diffusion affecting the various stake-holders associated with it.

## 1. INTRODUCTION
Internet has been a driving force towards the various technologies that have been developed since its inception. Arguably, one of the most discussed among all of them is *Cloud Computing*. Over the last few years, cloud computing paradigm has witnessed an enormous shift towards its adoption and it has become a trend in the information technology space as it promises significant cost reductions and new business potential to its users and providers [1]. The advantages of using cloud computing include: i) reduced hardware and maintenance cost, ii) accessibility around the globe, and iii) flexibility and highly automated processes wherein the customer need not worry about mundane concerns like software up-gradation [2, 3].

A plethora of definitions have been given explaining the cloud computing. Cloud computing is defined as a model for enabling ubiquitous, convenient, on-demand network access to a shared pool of configurable computing resources (e.g. networks, servers, storage devices and services) that can be rapidly provisioned and released with minimal management effort or service provider interaction [4]. In such an environment users need not own the infrastructure for various computing services. In fact, they can be accessed from any computer in any part of the world. This integrates features supporting high scalability and multi-tenancy, offering enhanced flexibility in comparison to the earlier existing computing methodologies. It can deploy, allocate or reallocate resources dynamically with an ability to continuously monitor their performance [4].

## 2. CLOUD TAXONOMY, CHARACTERISTICS AND BENEFITS
Cloud computing can be classified based on the services offered and deployment models. According to the different types of services offered, cloud computing can be considered to consist of three layers. Infrastructure as a Service (*IaaS*) is the lowest layer that provides basic infrastructure support service. Platform as a Service (*PaaS*) layer is the middle layer, which offers platform oriented services, besides providing the environment for hosting user's applications. Software as a Service *(SaaS)* is the topmost layer which features a complete application offered as service on demand [5, 6].

SaaS ensures that complete applications are hosted on the internet and users use them. The payment is made on a pay-per-use model. It eliminates the need to install and run the application on the customer's local computer, thus alleviating the customer's burden for software maintenance. In SaaS, there is the Divided Cloud and Convergence coherence mechanism whereby every data item has either the "Read Lock" or "Write Lock" [7]. Two types of servers are used by SaaS: the Main Consistence Server (MCS) and Domain Consistence Server (DCS). Cache coherence is achieved by the cooperation between MCS and DCS [8]. In SaaS, if the MCS is damaged, or compromised, the control over the cloud environment is lost. Hence securing the MCS is of great importance.

In the *Platform as a Service approach (PaaS)*, the offering also includes a software execution environment. For example, there could be a PaaS application server that enables the lone developer to deploy web-based applications without buying actual servers and setting them up. PaaS model aims to protect data, which is especially important in case of storage as a service. In case of congestion, there is the problem of outage from a cloud environment. Thus the need for security against outage is important to ensure load balanced service. The data needs to be encrypted when hosted on a platform for security reasons. Cloud computing architectures making use of multiple cryptographic techniques towards providing cryptographic cloud storage have been proposed in [9].

*Infrastructure as a Service (IaaS)* refers to the sharing of hardware resources for executing services, typically using virtualization technology. Potentially, with IaaS approach, multiple users use available resources. The resources can easily be scaled up depending on the demand from user and they are typically charged on a pay-per-use basis [10]. They are all virtual machines, which need to be managed. Thus a governance framework is required to control the creation and usage of virtual machines. This also helps to avoid uncontrolled access to user's sensitive information.

*Corresponding Author

Irrespective of the above mentioned service models, cloud services can be deployed in four ways depending upon the customers' requirements:

- *Public Cloud:* A cloud infrastructure is provided to many customers and is managed by a third party [11]. Multiple enterprises can work on the infrastructure provided, at the same time. Users can dynamically provision resources through the internet from an off-site service provider. Wastage of resources is checked as the users pay for whatever they use.

- *Private Cloud:* Cloud infrastructure, made available only to a specific customer and managed either by the organization itself or third party service provider [11]. This uses the concept of virtualization of machines, and is a proprietary network.

- *Community cloud:* Infrastructure shared by several organizations for a shared cause and may be managed by them or a third party service provider.

- *Hybrid Cloud:* A composition of two or more cloud deployment models, linked in a way that data transfer takes place between them without affecting each other.

Moreover, with the technological advancements, we can see derivative cloud deployment models emerging out of the various demands and the requirements of users. A virtual-private cloud is one such case wherein a public cloud is used in a private manner, connected to the internal resources of the customer's data-centre [12]. With the emergence of high-end network access technologies like 2G, 3G, Wi-Fi, Wi-Max etc. and feature phones, a new derivative of cloud computing has emerged. This is popularly referred to as "Mobile Cloud Computing (MCC)". It can be defined as a composition of mobile technology and cloud computing infrastructure where data and the related processing will happen in the cloud only with an exception that they can be accessed through a mobile device and hence termed as mobile cloud computing [13] as shown in Fig. 1. It is becoming a trend now-a-days and many organizations are keen to provide accessibility to their employees to access office network through a mobile device from anywhere.

Recent technical advancements including the emergence of HTML5 and various other browser development tools have only increased the market for mobile cloud-computing. An increasing trend towards the feature-phone adoption [13] has also ramped up the MCC market.

Cloud Computing distinguishes itself from other computing paradigms like grid computing, global computing, and internet computing in various aspects of on demand service provision, user centric interfaces, guaranteed QoS (Quality of Service), and autonomous system [14] etc. A few state of the art techniques that contribute to cloud computing are:

- *Virtualization:* It has been the underlying concept towards such a huge rise of cloud computing in the modern era. The term refers to providing an environment that is able to render all the services, supported by a hardware that can be observed on a personal computer, to the end users [15]. The three existing forms of virtualization categorized as: Server virtualization, Storage virtualization and Network virtualization, have inexorably led to the evolution of Cloud computing. For example, a number of underutilized physical servers may be consolidated within a smaller number of better utilized severs [16].

- *Web Service and SOA:* Web services provided services over the web using technologies like XML, Web Services Description Language (WSDL), Simple Object Access Protocol (SOAP), and Universal Description, Discovery, and Integration (UDDI). The service organisation inside a cloud is managed in the form of Service Oriented Architecture (SOA) and hence we can define SOA as something that makes use of multiple services to perform a specific task [17].

- *Application Programming Interface (API):* Without APIs it is hard to imagine the existence of cloud computing. The whole bunch of cloud services depend on APIs and allow deployment and configuration through them. Based on the API category used viz. control, data and application, different functions of APIs are invoked and services are rendered to the users accordingly.

- *Web 2.0 /Mash-up:* Web 2.0 has been defined as a technology that enables us to create web pages and allows the users to interact and collaborate as creators of user generated content in a virtual community [18, 19]. It enables the usage of World Wide Web technology towards a more creative and a collaborative platform [20]. Mash-up is a web application that combines data from more than one source into a single integrated storage tool [21].

These were the few technological advances that led to the emergence of Cloud Computing and enabled a lot of service providers to provide the customers a hassle free world of virtualization fulfilling all their demands. The prominent ones are: Amazon-EC2 [22, 23] (Elastic Compute Cloud), S3 [22] (Simple Storage Service), SQS (Simple Queue Service), CF (Cloud Front), SimpleDB, Google, Microsoft Windows-Azure [23], ProofPoint, RightScale, Salesforce.com, Workday, Sun Microsystems etc. and each of them are categorised either as one of the three main classifications based on the cloud structure they provide: private, public and hybrid cloud. Each of the above mentioned cloud structure has its own limitations and benefits.

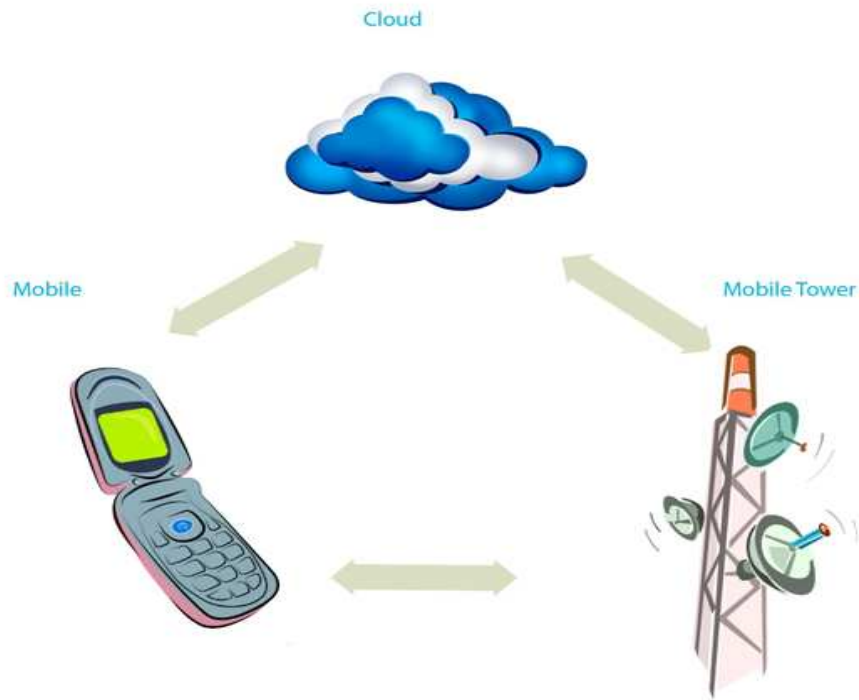

**Fig 1: A Mobile Cloud Computing Scenario**

The enormous growth in this field has changed the way computing world is perceived. The IT sector has witnessed the change in the way situations are handled. However, there are issues that still persist and have become even more compelling now. The amount of significant resources available at very low price is acting as a catalyst for distributed attacks on confidential information.

With a substantial increase in the number of Cloud Computing deployments, the issues related to security and privacy have become more sophisticated and more distributed in the sense that the user section for such services is growing by leaps and bounds [24, 25]. With an increase in on-demand application usage, the possibility of cyber attacks also increases. Individual users have to frequently provide online information about their identification, and this could be used by attackers for identity theft. In order to maintain various security and privacy issues like: confidentiality, operational integrity, disaster recovery and identity management, following schemes should be deployed at least to ensure data security [26] to some extent:

- An encryption scheme to ensure data security in a highly interfering environment, maintaining security standards against popular threats to data storage security.
- The Service Providers should be given limited access to the data, just to manage it without being able to see what exactly the data is.
- Stringent access controls to prevent unauthorized and illegal access to the servers controlling the network.
- Data backup and redundant data storage to ensure seamless data retrieval in case of infrastructure failure like the recent breakdown issues with the Amazon cloud.
- Distributed identity management and user security is to be maintained by using either Lightweight Directory Access Protocol (LDAP), or published APIs (Application Programming Interfaces) to connect into identity systems.

An important aspect of cloud computing is that it does give rise to a number of security threats from the perspective of data security for a couple of reasons. Firstly, the traditional techniques cannot be adopted as these have become quite obsolete with respect to the ever evolving security threats and also to avoid data loss in a cloud computing environment. The second issue is that the data stored in the cloud is accessed a large number of times and is often subjected to different types of changes. This may comprise of bank accounts, passwords and highly confidential files, not to be read by someone else apart from the owner. Hence, even a small error may result in loss of data security.

This paper is aimed at developing an understanding of the manifold security threats that hamper the security and privacy of a user. Characteristics of a secure cloud infrastructure (public or private) will be discussed as also its challenges and the ways to solve them.

## 3. OBSTACLES AND OPPORTUNITIES FOR CLOUD COMPUTING

In spite of being a buzzword, there are certain aspects associated with Cloud Computing as a result of which many organizations are still not confident about moving into the cloud. Certain loopholes in its architecture have made cloud computing vulnerable to various security and privacy threats [27]. A few issues limiting the boundaries of this transformational concept are:

## 3.1 Privacy and Security

The fundamental factor defining the success of any new computing technology is the level of security it provides [28, 29, 30]. Whether the data residing in the cloud is secure to a level so as to avoid any sort of security breach or is it more secure to store the data away from cloud in our own personal computers or hard drives? At-least we can access our hard drives and systems whenever we wish to, but cloud servers could potentially reside anywhere in the world and any sort of internet breakdown can deny us access to the data stored in the cloud. The cloud service providers insist that their servers and the data stored in them is sufficiently protected from any sort of invasion and theft. Such companies argue that the data on their servers is inherently more secure than data residing on a myriad of personal computers and laptops. However, it is also a part of cloud architecture, that the client data will be distributed over these individual computers regardless of where the base repository of data is ultimately located. There have been instances when their security has been invaded and the whole system has been down for hours. At-least half a dozen of security breaches occurred last year bringing out the fundamental limitations of the security model of major Cloud Service Providers (CSP). With respect to cloud computing environment, privacy is defined as "the ability of an entity to control what information it reveals about itself to the cloud/cloud SP, and the ability to control who can access that information". R. Gellman discusses the standards for collection, maintenance and disclosure of *personality identifiable information* in [24]. Information requiring privacy and the various privacy challenges need the specific steps to be taken in order to ensure privacy in the cloud as discussed in [31 , 32].

In case of a public-cloud computing scenario, we have multiple security issues that need to be addressed in comparison to a private cloud computing scenario. A public cloud acts as a host of a number of virtual machines, virtual machine monitors, and supporting middleware [33] etc. The security of the cloud depends on the behaviour of these objects as well as on the interactions between them. Moreover, in a public cloud enabling a shared multi-tenant environment, as the number of users increase, security risks get more intensified and diverse. It is necessary to identify the attack surfaces which are prone to security attacks and mechanisms ensuring successful client-side and server-side protection [34]. Because of the multifarious security issues in a public cloud, adopting a private cloud solution is more secure with an option to move to a public cloud in future, if needed [35].

Emergence of cloud computing owes significantly to mashup. A mashup is an application that combines data, or functionality from multiple web sources and creates new services using these. As these involve usage of multiple sub-applications or elements towards a specific application, the security challenges are diverse and intense. Based on this idea, various security architectures such as: a secure component model addressing the problem of securing mash-up applications and an entropy based security framework for cloud oriented service mash-ups have been proposed in [36, 66]. Also, privacy needs to be maintained as there are high chances of an eavesdropper to be able to sneak in.

## 3.2 Performance Unpredictability, Latency and Reliability

It has been observed that virtual machines can share CPUs and main memory in a much better way in comparison to the network and disk I/O. Different EC2 instances vary more in their I/O performance than main memory performance [37]. One of the ways to improve I/O performance is to improve architecture and operating systems to efficiently virtualize interrupts and I/O channels. Another possibility is to make use of flash memory which is a type of semiconductor memory that preserves information even when powered off and since it has no moving parts, it is much faster to access and uses comparatively less energy. Flash memory can sustain many more I/O operations than disks, so multiple virtual machines with large number of I/O operations would coexist better on the same physical computer [37].

Latency [38, 39] has always been an issue in cloud computing with data expected to flow around different clouds. The other factors that add to the latency are: encryption and decryption of the data when it moves around unreliable and public networks, congestion, packet loss and windowing. Congestion adds to the latency when the traffic flow through the network is high and there are many requests (could be of same priority) that need to be executed at the same time. Windowing is another message passing technique whereby the receiver has to send a message to the sender that it has received the earlier sent packet and hence this additional traffic adds to the network latency. Moreover, the performance of the system is also a factor that should be taken into account. Sometimes the cloud service providers run short of capacity either by allowing access to too many virtual machines or reaching upper throughput thresholds on their Internet links because of high demand arising from the customer community. This affects the system performance and adds to the latency of the system.

## 3.3 Portability and Interoperability

Organizations may need to change the cloud providers and there have been cases when companies are unable to move their data and applications to another cloud platform that they would prefer over the existing one. Such a scenario is termed as Lock-in and it refers to the challenges faced by a cloud customer trying to migrate from one cloud provider to another. More often, it has been seen that changing a cloud provider involves multiple risks and may lead to system breakdown if not executed properly. Nature of Lock-in and associated issues are very much dependent on the cloud type being used [40]. In case of a SaaS offering, an application is used by the customer provided by the cloud provider. While migrating between the cloud providers, there may be instances when the data to be moved does not really fit the data format as required in the new application. This will require extra effort to be put in to make sure that the data is arranged in a format that matches the new application ensuring no data loss in the process. Additional steps such as: performing regular data extraction and back-up to a format that is usable even without the SaaS application, understanding how the application has been developed and monitored, and the major interfaces and their integration between the platforms need to be taken care of.

PaaS lock-in can be observed in cases where the language used to develop an application on a platform is not supported on the platform to be migrated to. It is more visible at API level as different providers offer different APIs. PaaS lock-in

can be avoided if the following points are considered and addressed:

- Cloud offering with an open architecture and standard syntax should be supported.
- Understand the application components and modules specific to the PaaS provider and how the basic services like monitoring, logging etc. are performed.
- Understand the control functions specific to the cloud provider and their counterparts on an Open platform.

IaaS lock-in depends on the infrastructure services being used. The most obvious form of IaaS lock-in can be observed in the form of data lock-in. With more and more data pushed to the cloud, data lock-in increases unless the cloud provider ensures data portability. Understanding how the virtual machine images are maintained and eliminating any provider specific dependency for a virtual machine environment will serve at the time of transition from one IaaS platform to other. Identifying the hardware dependencies will minimize the issues at the time of migration. In order to avoid this lock-ins, the customer should be clear of the choices available in the market and the extent to which they match up to its business, operational and technical requirements.

Also, some companies use different cloud platforms for different applications based on their requirements and the services provided by the cloud service providers (CSPs). In some cases, different cloud platforms are used for a particular application or different cloud platforms have to interact with each other for completing a particular task. The internal infrastructure of the organization is needed to maintain a balance to handle the interoperability between different cloud platforms [41]. The risk of outsourced services going out of control is high in a hybrid, public and private cloud environment. All data has to be encrypted for proper security, and key management becomes a difficult task in such situations [42]. The users have actually no idea of where their information is stored [43]. Normally, a user's data is stored in a shared environment, along-with other user's data. The issue of inter-security handling becomes important in such cases. A cloud security management model is discussed in [42] to serve as a standard for designing cloud security management tools. The model uses four interoperating layers for managing the cloud security.

Thus we see that although the buzz of cloud computing prevails everywhere because of the multi-fold features and facilities provided by it, there are still issues that need to be solved in order to reach the landmarks set by it.

### 3.4 Data Breach through Fibre Optic Networks

It has been noticed that the security risks for the data in transit has increased over the last few years. Data transitioning is quite normal now-a-days and it may include multiple data-centres and other cloud deployment models such as public or private cloud. Security of the data leaving a data-centre to another data-centre is a major concern as it has been breached quite a number of times in the recent times.

This data transfer is done over a network of fibre-optic cables which were considered to be a safe mode of data-transfer, until recently an illegal fibre eavesdropping device in Telco Verizon's optical network placed at a mutual fund company was discovered by US Security forces [44]. There are devices that can tap the data flow without even disturbing it and accessing fibre, through which data is being transferred. They are generally laid underground and hence it is difficult to access these fibre-optic cables. And hence it becomes quite important to ensure data security over the transitioning networks.

### 3.5 Data Storage over IP Networks

Online data storage is becoming quite popular now-a-days and it has been observed that majority of enterprise storage will be networked in the coming years, as it allows enterprises to maintain huge chunks of data without setting up the required architecture. Although there are many advantages of having online data storage, there are security threats that could cause data leakage or data unavailability at crucial hour. Such issues are observed more frequently in the case of dynamic data that keeps flowing within the cloud in comparison to static data. Depending upon the various levels of operations and storage provided, these networked devices are categorized into SAN (Storage area network) and NAS (network-attached storage) and since these storage networks reside on various servers, there are multiple threats associated with them. Various threat zones that may affect and cause the vulnerability of a storage network have been discussed in [45].

Besides these, from a Mobile Cloud Computing (MCC) perspective, unlike cloud computing there are several additional challenges that need to be addressed to enable MCC reach its maximum potential:

- *Network accessibility:* Internet has been the major factor towards the cloud computing evolution and without having the network (Internet) access it will not be possible to access the mobile cloud limiting the available applications that can be used.

- *Data Latency:* Data transfer in a wireless network is not as continuous and consistent as it is in case of a dedicated wired LAN. And this inconsistency is largely responsible for longer time intervals for data transfer at times. Also, the distance from the source adds up to the longer time intervals observed in case of data transfer and other network related activities because of an increase in the number of intermediate network components.

- *Dynamic Network monitoring and Scalability:* Applications running on mobiles in a mobile cloud computing platform should be intelligent enough to adapt to the varying network capacities and also these should be accessible through different platforms without suffering any data loss. Sometimes, a user while working on a smart phone may need to move on to a feature phone and when he accesses the application through a smart phone; he should not encounter any data loss.

- *Confidentiality of mobile cloud-based data sharing*: The confidential data on mobile phones using cloud-based mobile device support might become public due to a hacked cloud. The root-level access to cloud services and information can be easily accessed from a stolen mobile device. If the stolen device belongs to a system administrator, they may even provide direct and automated access to highly confidential information.

- *Better access control and identity management:* Cloud computing involves virtualization, and hence the need for user authentication and control across the clouds is high. The existing solutions are not able to handle the case of multiple clouds. Since data belonging to multiple users may be stored in a single hypervisor, specific segmentation measures are needed to overcome the potential weakness and flaws in hypervisor platform.

Security challenges in a mobile cloud computing environment are slightly different as compared to the above mentioned network related challenges. With applications lying in a cloud, it is possible for the hackers to corrupt an application and gain access to a mobile device while accessing that application. In order to avoid these, strong virus-scanning and malware protection software need to be installed to avoid any type of virus/malware check into the mobile system. Besides, by embedding device identity protection, like allowing access to the authorized user based on some form of identity check feature, unauthorized accesses can be blocked.

Two types of services, have been defined in [46], namely (i) critical security service, and (ii) normal security service. The resource in a cloud has to be properly partitioned according to different user's requests. The maximal system rewards and system service overheads are considered for the security service. Hence, we see that although mobile cloud computing is still in its nascent state, there are various security issues, that plague cloud computing and its derivatives.

## 4. DATA STORAGE AND SECURITY IN THE CLOUD

Many cloud service providers provide storage as a form of service. They take the data from the users and store them on large data centres, hence providing users a means of storage. In spite of claims by the cloud service providers about the safety of the data stored in the cloud there have been cases when the data stored in these clouds have been modified or lost due to some security breach or some human error. Attack vectors in a cloud storage platform have been discussed and how the same platform is exploited to hide files with unlimited storage in [47]. In [47], authors have studied the storage mechanism of Dropbox (a file storage solution in the cloud) and carried three types of attack viz. Hash Value manipulation attack, stolen host id attack and direct download attack. Once the host id is known, the attacker can upload and link arbitrary files to the victim's Dropbox account.

Various cloud service providers adopt different technologies to safeguard the data stored in their cloud. But the question is: Is the data stored in these clouds really secure? The virtualized nature of cloud storage makes the traditional mechanisms unsuitable for handling the security issues [23]. These service providers use different encryption techniques such as: public key encryption and private key encryption to secure the data stored in the cloud. A similar technique providing data storage security, utilizing the homomorphic token with distributed verification of erasure-coded data has been discussed in [48]. Trust based methods are useful in establishing relationships in a distributed environment. A domain based trust-model has been proposed in [49] to handle security and interoperability in cross clouds. Every domain has a special agent for trust management. It proposes different trust mechanisms for users and service providers.

The following aspects of data security should be taken care while moving into a cloud:

1. Data-in-transit
2. Data-at-rest
3. Data Lineage
4. Data Remanence
5. Data Provenance

In case of data-in-transit, the biggest risk is associated with the encryption technology that is being used, whether it is up-to-date with the present day security threats and makes use of a protocol that provides confidentiality as well as integrity to the data-in-transit. Simply going for an encryption technology does not serve the purpose. In addition to using an encryption – decryption algorithm for secure data transfer, data can be broken into packets and then transferred through disjoint paths to the receiver. It reduces the chances of all the packets being captured by an adversary. And the data cannot be known until all the packets are coupled together in a particular manner. A similar approach has been discussed in [50, 51].

Managing data at rest in an IaaS scenario is more feasible in comparison to managing the same over a SaaS and PaaS platform because of restricted rights over the data. In a SaaS and PaaS platform, data is generally commingled with other users' data. There have been cases wherein even after implementing data tagging to prevent unauthorized access, it was possible to access data through exploitation of application vulnerability [25]. The main issue with data-at-rest in the cloud is loss of control, even a non-authorized user/party may have access to the data (it is not supposed to access) in a shared environment. However, now-a-days, storage devices with in-built encryption techniques are available which are resilient to unauthorized access to certain extent. Even in such a case, nothing can be done in case the encryption and decryption keys are accessible to the malicious user. A lockbox approach wherein the actual keys are stored in a lockbox and there is a separate key to access that lockbox is useful in the above mentioned case. In such a scenario, a user will be provided a key based on identity management technique corresponding to the COI (community of interest) he belongs to, to access the lockbox. Whenever the user wants to access the data, he needs to acquire the COI key to the lockbox and then the user gets appropriate access to the relevant data [9]. Homomorphic encryption techniques, which are capable of processing the encrypted data and then bringing back the data into its original form, are also providing better means to secure the data-at-rest. A simple technique for securing data at rest in a cloud computing environment has been mentioned in [52]. This technique makes use of public encryption technique.

Tracing the data path is known as data lineage and it is important for auditing purpose in the cloud. Providing data lineage is a challenging task in a cloud computing environment and more so in a public cloud. Since the data flow is no longer linear in a virtualized environment within the cloud, it complicates the process of mapping the data flow to ensure integrity of the data. Proving data provenance is yet another challenging task in a cloud computing environment. Data provenance refers to maintaining the integrity of the data, ensuring that it is computationally correct. Taxonomy of provenance techniques and various data provenance techniques have been discussed in [53].

Another major issue that is mostly neglected is of Data-Remanence. It refers to the data left out in case of data transfer or data removal. It causes minimal security threats in private cloud computing offerings, however severe security issues may emerge out in case of public cloud offerings as a result of data-remanence [54, 56].

Various cases of cloud security breach came into light in recent past. Cloud based email marketing services company, Epsilon, suffered a data breach, due to which a large section of its customers including JP Morgan Chase, Citibank, Barclays Bank, hotel chains such as Marriott and Hilton, and big retailers such as Best Buy and Walgreens were affected heavily and huge chunk of customer data was exposed to the hackers which includes customer email ids and bank account details [55].

A similar incident happened with Amazon causing the disruption of its EC2 services. Popular sites like: Quora, Four-Square and Reditt were the main sufferers [57]. The above mentioned events depict the vulnerability of the cloud services.

Another important aspect is that the known and popular domains have been used to launch malicious software or hack into companies' secure database. A similar issue happened with Amazon's S3 platform and the hackers were able to launch corrupted codes using a trusted domain [58]. Hence the question that arises now is who to be provided the "trusted" tag. It established that Amazon was prone to side-channel attacks, and a malicious virtual machine, occupying the same server as the target, could easily gain access to the confidential data [59]. The question is: should any such security policy be in place for these trusted users as well?

An incident related to the data loss occurred, sometime back, with the online storage service provider "Media max" (also known as "The Linkup") when due to system administration error; active customer data was deleted, leading to huge data loss [60]. SLA (Service Level Agreement) with the Cloud Service providers should contain all the points that may cause data loss either due to some human or system generated error. Hence, it must be ensured that redundant copies of the user data should be stored in order to handle any sort of adverse situation leading to data loss.

Virtualization in general increases the security of a cloud environment. With virtualization, a single machine can be divided into many virtual machines, thus providing better data isolation and safety against denial of service attacks [68]. The VMs (Virtual Machine) provide a security test-bed for execution of untested code from un-trusted users. A hierarchical reputation system has been proposed in the paper [61] for managing trust in a cloud environment.

## 5. THREATS TO SECURITY IN CLOUD COMPUTING

The chief concern in cloud environments is to provide security around multi-tenancy and isolation, giving customers more comfort besides "trust us" idea of clouds [62]. There has been survey works reported, which classify security threats in cloud based on the nature of the service delivery models of a cloud computing system [63]. However, security requires a holistic approach. Service delivery model is one of many aspects that need to be considered for a comprehensive survey on cloud security. Security at different levels such as Network level, Host level and Application level is necessary to keep the cloud up and running continuously and the same has been discussed in [64] for Amazon EC2 cloud. In accordance with these different levels, various types of security breaches may occur which have been classified in this section.

### 5.1 Basic Security

Web 2.0, a key technology towards enabling the use of Software as a Service (SaaS) relieves the users from tasks like maintenance and installation of software. It has been used widely all around. As the user community using Web 2.0 is increasing by leaps and bounds, the security has become more important than ever for such environment [65, 67].

*SQL injection attacks*, are the one in which a malicious code is inserted into a standard SQL code. Thus the attackers gain unauthorized access to a database and are able to access sensitive information [68]. Sometimes the hacker's input data is misunderstood by the web-site as the user data and allows it to be accessed by the SQL server and this lets the attacker to have know-how of the functioning of the website and make changes into that. Various techniques like: avoiding the usage of dynamically generated SQL in the code, using filtering techniques to sanitize the user input etc. are used to check the SQL injection attacks. A proxy based architecture towards preventing SQL Injection attacks which dynamically detects and extracts users' inputs for suspected SQL control sequences has been proposed in [69].

*Cross Site Scripting (XSS) attacks,* which inject malicious scripts into Web contents have become quite popular since the inception of Web 2.0. There are two methods for injecting the malicious code into the web-page displayed to the user: Stored XSS and Reflected XSS. In a Stored XSS, the malicious code is permanently stored into a resource managed by the web application and the actual attack is carried out when the victim requests a dynamic page that is constructed from the contents of this resource [70]. However, in case of a Reflected XSS, the attack script is not permanently stored; in fact it is immediately reflected back to the user [70].

Based on the type of services provided, a website can be classified as static or dynamic. Static websites do not suffer from the security threats which the dynamic websites do because of their dynamism in providing multi-fold services to the users. As a result, these dynamic websites get victimized by XSS attacks. It has been observed quite often that amidst working on the internet or surfing, some web-pages or pop-ups open up with the request of being clicked away to view the content contained in them. More often either unknowingly (about the possible hazards) or out of curiosity users click on these hazardous links and thus the intruding third party gets control over the user's private information or hack their accounts after having known the information available to them. Various techniques like: Active Content Filtering, Content Based Data Leakage Prevention Technology, Web Application Vulnerability Detection Technology has already been proposed to prevent XSS attacks [71]. These technologies adopt various methodologies to detect security flaws and fix them. A blueprint based approach that minimizes the dependency on web browsers towards identifying untrusted content over the network has been proposed in [72].

Another class of attacks, quite popular to SaaS, are termed as *Man in the Middle attacks (MITM)*. In such an attack, an entity tries to intrude in an ongoing conversation between a sender and a client to inject false information and to have knowledge of the important data transferred between them. Various tools implementing strong encryption technologies like: Dsniff, Cain, Ettercap, Wsniff, Airjack etc. have been developed in order to provide safeguard against them. A detailed study towards preventing man in the middle attacks has been presented in [73].

A few of the important points like: evaluating software as a service security, separate endpoint and server security processes, evaluating virtualization at the end-point have been mentioned by Eric Ogren, in an article at Security.com to tackle traditional security flaws [74].

Hence, security at different levels is necessary in order to ensure proper implementation of cloud computing environment such as: server access security, internet access security, database access security, data privacy security and program access security. In addition, we need to ensure data security at network layer, and data security at physical and application layer to maintain a secure cloud.

## 5.2 Network Level Security

Networks are classified into different types like: shared and non-shared, public or private, small area or large area networks and each of them have a number of security threats to deal with. While considering the network level security, it is important to distinguish between public and private clouds. There is less vulnerability in a private cloud in comparison to public cloud. Almost all the organizations have got a private network in place and hence the network topology for a private cloud gets defined. And in most of the cases, the security practices implemented (in the organization's private network) apply to the private cloud too. However, in case of a public cloud implementation, network topology might need to be changed in order to implement the security features and the following points need to be addressed as part of public cloud implementation:

- Confidentiality and Integrity of the data-in-transit needs to be ensured while adopting a public cloud architecture.

- Ensuring proper access controls within the cloud.
    o Migrating to a cloud exposes the resources to Internet and the data which has been hosted over a private network till now, becomes accessible over the internet. This also increases the chances of data leakage or a security breach which should be taken care of.
    o It may happen that the security policies implemented inside the cloud are not up to date and as a result other parties within the cloud are able to access data belonging to some other customer.
- The trusted encryption schemes and tokenization models need to be changed to enhance the security in a public cloud.

We can now see the reasons because of which organizations are not moving their sensitive data to public clouds and instead relying on private cloud. In addition to the concerns mentioned above, issues associated with network level security comprise of: DNS attacks, Sniffer attacks, issue of reused IP address, Denial of Service (DoS) and Distributed Denial of Service attacks (DDoS) etc. [75].

### *5.2.1 DNS Attacks*
Domain Name Server (DNS) performs the translation of a domain name to an IP address since the domain names are much easier to remember. Hence, the DNS servers are needed. But there are cases when having called the server by name, the user has been routed to some other malicious cloud instead of the one he asked for and hence using IP address is not always feasible. Although using DNS security measures like: Domain Name System Security Extensions (DNSSEC) reduces the effects of DNS threats but still there are cases when these security measures prove to be inadequate when the path between a sender and a receiver gets rerouted through some malicious connection. It may happen that even after all the DNS security measures are taken, the route selected between the sender and receiver cause security problems [76].

### *5.2.2 Sniffer Attacks*
These types of attacks are launched by applications which can capture packets flowing in a network and if the data that is being transferred through these packets is not encrypted, it can be read. There are chances that vital information flowing across the network can be traced or captured. A sniffer program, through the NIC (Network Interface Card) ensures that the data/traffic linked to other systems on the network also gets recorded. It can be achieved by placing the NIC in promiscuous mode and in promiscuous mode it can track all data, flowing on the same network. A malicious sniffing detection platform based on ARP (address resolution protocol) and RTT (round trip time) can be used to detect a sniffing system running on a network [77].

### *5.2.3 Issue of Reused IP Addresses*
Each node of a network is provided an IP address and the number of IP addresses that can be assigned is limited. A large number of cases related to re-used IP-address issue have been observed lately. When a particular user moves out of a network, then the IP-address associated with him (earlier) is assigned to a new user. This sometimes risks the security of the new user as there is a certain time lag between the change of an IP address in DNS and the clearing of that address in DNS caches [25]. And hence, we can say that sometimes though the old IP address is being assigned to a new user still the chances of accessing the data by some other user is not negligible as the address still exists in the DNS cache and the data belonging to a particular user may become accessible to some other user violating the privacy of the earlier user.

### *5.2.4 BGP Prefix Hijacking*
Prefix hijacking is a type of network attack in which a wrong announcement related to the IP addresses associated with an Autonomous system (AS) is made. Hence, malicious parties get access to the untraceable IP addresses. On the internet, IP space is associated in blocks and remains under the control of ASs. An autonomous system can broadcast information of an IP contained in its regime to all its neighbours.

These ASs communicate using the Border Gateway Protocol (BGP) model. Sometimes, due to some error, a faulty AS may broadcast wrongly about the IPs associated with it. In such case, the actual traffic gets routed to some IP other than the intended one. Hence, data is leaked or reaches to some other unintended destination. A security system for autonomous systems has been explained in [78].

## 5.3 Application Level Security
Application level security refers to the usage of software and hardware resources to provide security to applications such that the attackers are not able to get control over these applications and make desirable changes to their format. Now a days, attacks are launched, being disguised as a trusted user and the system considering them as a trusted user, allows full access to the attacking party and gets victimized. The reason behind this is that the outdated network level security policies allow only the authorized users to access the specific IP address. With the technological advancement, these security policies have become obsolete as there have been instances when the system's security has been breached, having accessed the system in the disguise of a trusted user. With the recent technological advancements, it is quite possible to imitate a trusted user and corrupt entire data without being noticed.

Hence, it is essential to install higher level of security checks to minimize these risks. The traditional methods to deal with increased security issues have been to develop a task oriented ASIC device which can handle a specific task, providing greater levels of security with high performance [79]. But with application-level threats being dynamic and adaptable to the security checks in place, these closed systems have been observed to be slow in comparison to the open ended systems.

The capabilities of a closed system as well as the adaptability of an open ended system have been incorporated to develop the security platforms based on Check Point Open Performance Architecture using Quad Core Intel Xeon Processors [79]. Even in the virtual environment, companies like VMware etc. are using Intel Virtualization technology for better performance and security base. It has been observed that more often websites are secured at the network level and have strong security measures but there may be security loopholes at the application level which may allow information access to unauthorized users. The threats to application level security include XSS attacks, Cookie Poisoning, Hidden field manipulation, SQL injection attacks, DoS attacks, Backdoor and Debug Options, CAPTCHA Breaking etc. resulting from the unauthorized usage of the applications.

### 5.3.1 Security Concerns with the Hypervisor

Cloud Computing rests mainly on the concept of virtualization. In a virtualized world, hypervisor is defined as a controller popularly known as virtual machine manager (VMM) that allows multiple operating systems to be run on a system at a time, providing the resources to each operating system such that they do not interfere with each other.

As the number of operating systems running on a hardware unit increase, the security issues concerned with those new operating systems also need to be considered. Because multiple operating systems would be running on a single hardware platform, it is not possible to keep track of all such systems and hence maintaining the security of the operating systems is difficult. It may happen that a guest system tries to run a malicious code on the host system and bring the system down or take full control of the system and block access to other guest operating systems [80].

It cannot be denied that there are risks associated with sharing the same physical infrastructure between a set of multiple users, even one being malicious can cause threats to the others using the same infrastructure [81], and hence security with respect to hypervisor is of great concern as all the guest systems are controlled by it. If a hacker is able to get control over the hypervisor, he can make changes to any of the guest operating systems and get control over all the data passing through the hypervisor.

Various types of attacks can be launched by targeting different components of the hypervisor [82]. Based on the understanding of how the various components in the hypervisor architecture behave, an advanced cloud protections system can be developed by monitoring the activities of the guest VMs (Virtual Machines) and inter-communication among the various infrastructure components [83, 84].

### 5.3.2 Denial of Service Attacks

A DoS attack is an attempt to make the services assigned to the authorized users unavailable. In such an attack, the server providing the service is flooded by a large number of requests and hence the service becomes unavailable to the authorized user. Sometimes, when we try to access a site we see that due to overloading of the server with the requests to access the site, we are unable to access the site and observe an error. This happens when the number of requests that can be handled by a server exceeds its capacity. The occurrence of a DoS attack increases bandwidth consumption besides causing congestion, making certain parts of the clouds inaccessible to the users. Usage of an Intrusion Detection System (IDS) is the most popular method of defence against this type of attacks [85]. A defence federation is used in [31] for guarding against such attacks. Each cloud is loaded with separate IDS. The different intrusion detection systems work on the basis of information exchange. In case a specific cloud is under attack, the co-operative IDS alerts the whole system. A decision on trustworthiness of a cloud is taken by voting, and the overall system performance is not hampered.

### 5.3.3 Cookie Poisoning

It involves changing or modifying the contents of cookie to have an unauthorized access to an application or to a web-page. Cookies basically contain the user's identity related credentials and once these cookies are accessible, the content of these cookies can be forged to impersonate an authorized user. This can be avoided either by performing regular cookie cleanup or implementing an encryption scheme for the cookie data [71].

### 5.3.4 Hidden Field Manipulation

While accessing a web-page, there are certain fields that are hidden and contain the page related information and basically used by developers. However, these fields are highly prone to attacks by hackers as they can be modified easily and posted on the web-page. This may result in severe security violations [86].

### 5.3.5 Backdoor and Debug Options

A common practice by the developers is to enable the debug option while publishing a web-site. This enables them to make developmental changes in the code and get them implemented in the web-site. Since these debug options facilitate back-end entry to the developers, and sometimes these debug options are left enabled unnoticed, this may provide an easy entry to a hacker into the web-site that let him make changes at the web-site level [87].

### 5.3.6 Distributed Denial of Service Attacks

DDoS may be called an advanced version of DoS in terms of denying the important services running on a server by flooding the destination sever with large numbers of packets such that the target server is not able to handle it. In DDoS the attack is relayed from different dynamic networks which have already been compromised unlike the DoS attack. The attackers have the power to control the flow of information by allowing some information available at certain times. Thus the amount and type of information available for public usage is clearly under the control of the attacker [87].

The DDoS attack is run by three functional units: A Master, A Slave and A Victim. Master being the attack launcher is behind all these attacks causing DDoS, Slave is the network which acts like a launch pad for the Master. It provides the platform to the Master to launch the attack on the Victim. Hence it is also called as co-ordinated attack.

Basically a DDoS attack is operational in two stages: the first one being Intrusion phase where the Master tries to compromise less important machines to support in flooding the more important one. The next one is installing DDoS tools and attacking the victim server or machine. Hence, a DDoS attack results in making the service unavailable to the authorized user similar to the way it is done in a DoS attack but different in the way it is launched. A similar case of Distributed Denial of Service attack was experienced with CNN news channel website leaving most of its users unable to access the site for a period of three hours [88].

In general, the approaches used to fight the DDoS attack involve extensive modification of the underlying network. These modifications often become costly for the users. [87] proposed a swarm based logic for guarding against the DDoS attack. This logic provides a transparent transport layer, through which the common protocols such as HTTP, SMTP, etc. can pass easily. The use of IDS in the virtual machine is proposed in [16] to protect the cloud from DDoS attacks. A SNORT like intrusion detection mechanism is loaded onto the virtual machine for sniffing all traffics, either incoming, or outgoing. Another method commonly used to guard against DDoS is to have intrusion detection systems on all the physical machines which contain the user's virtual machines [89]. This scheme had been shown to perform reasonably well in a Eucalyptus [90] cloud.

### 5.3.7 CAPTCHA Breaking

CAPTCHAs were developed in order to prevent the usage of internet resources by bots or computers. They are used to prevent spam and overexploitation of network resources by bots. Even multiple web-site registrations, dictionary attacks etc. by an automated program are prevented using a CAPTCHA.

But recently, it has been found that the spammers are able to break the CAPTCHA [91], provided by the Hotmail and G-mail service providers. They make use of the audio system able to read the CAPTCHA characters for the visually impaired users and use speech to text conversion software to defeat the test. In yet another instant of CAPTCHA Breaking, it was found that the net users are provided some form of motivation towards solving these CAPTCHA's by the automated systems and thus CAPTCHA Breaking takes place. Integration of multiple authentication techniques along with CAPTCHA identification (as adopted by companies like Facebook, Google etc.) may be a suitable option against CAPTCHA breaking. Various techniques such as: implementing letter overlap, variable fonts of the letters used to design a CAPTCHA, increasing the string length and using a perturbative background can be used to avoid CAPTCHA breaking [92].

A safe CAPTCHA design framework based on the problems of multiple moving object recognition in complex background has been presented in [93]. Single frame zero knowledge CAPTCHA design principles have been proposed, which will be able to resist any attack method of static optical character recognition (OCR). Such a design to create CAPTCHAs will be resistant to attack methods launched by intercepting picture to identify or intercepting each video frame to recognize the CAPTCHA separately.

### 5.3.8 Dictionary Attack

Data security in a cloud computing environment can be compromised by carrying out a dictionary or brute force attack. In a dictionary attack, the intruder makes use of all the possible word combinations which could have been successfully used to decrypt the data residing in/flowing over the network. They can be avoided by making use of a challenge-response system as explained in [94]. In this protocol, the client is presented a challenge whenever it tries to access a network. It is then required to compute the response to the same and reply back to the server in order to be able to access the network. Response computation is a time consuming process thus avoiding the users to be able to launch brute force or dictionary attacks in a short period of time and hence ensuring security against the same.

### 5.3.9 Google Hacking

Google has emerged as the best option for finding details regarding anything on the internet. Google hacking refers to using Google search engine to find sensitive information that a hacker can use to his benefit while hacking a user's account. Generally, hackers try to find out the security loopholes by probing out on Google about the system they wish to hack. After having gathered the necessary information, they carry out the hacking of the concerned system. In some cases, a hacker is not sure of the target. Instead he tries to discover the target, using Google, based on the loophole he wishes to hack a system upon. The hacker then searches all the possible systems with such a loophole and finds out those having the loopholes he wishes to hack upon. A Google hacking event was observed recently when login details of various Gmail users were stolen by a group of hackers [95].

These have been some of the security threats that can be launched at the application level and cause a system downtime disabling the application access even to the authorized users. In order to avoid these threats, application security should be assessed at the various levels of the three service delivery models in cloud: IaaS, PaaS and SaaS. In case of an IaaS delivery model, cloud providers are mostly not concerned with the security policies applied by the customer and the application's management. The whole application runs on the customer's server on the cloud provider's infrastructure and managed by them and hence responsible for securing the application. The following points should be taken care of while designing the application:

- Standard security measures must be implemented to safeguard against the common vulnerabilities associated with the web.

- Custom implementation of authorization and authentication schemes should not be implemented unless they are tested properly.

- Back up policies such as Continuous Data Protection (CDP) should be implemented in order to avoid issues with data recovery in case of a sudden attack [96].

Additionally, they should be aware if the virtual network infrastructure used by the cloud provider is secured and the various security procedures implemented to ensure the same [25]. Security challenges for IaaS Cloud Computing and multiple levels of security as operational in Amazon EC2 cloud have been discussed in [64]. It discusses identity/access management and multifactor authentication techniques in Amazon Web Service (AWS) cloud.

PaaS service providers are responsible for maintaining the security of the platform an application is built upon and the following aspects should be considered to assess the security policy of the service provider:

- How the different applications on PaaS platform are isolated from each other and whether the data

- belonging to one customer is inaccessible to any other customer or not (in case of public cloud).
- Does the service provider keep checking and updating its security policies at regular intervals and ensure that the new security policies are implemented.

The above mentioned concerns apply to a SaaS scenario as well, as the security control lies with the provider instead of the customer.

# 6. SECURITY ISSUES IN THE CLOUD DEPLOYMENT MODELS

Each of the three ways in which cloud services can be deployed has its own advantages and limitations. And from the security perspective, all the three have got certain areas that need to be addressed with a specific strategy to avoid them.

## 6.1 Security issues in a public cloud

In a public cloud, there exist many customers on a shared platform and infrastructure security is provided by the service provider. A few of the key security issues in a public cloud include:

1) The three basic requirements of security: confidentiality, integrity and availability are required to protect data throughout its lifecycle. Data must be protected during the various stages of creation, sharing, archiving, processing etc. However, situations become more complicated in case of a public cloud where we do not have any control over the service provider's security practices [97].

2) In case of a public cloud, the same infrastructure is shared between multiple tenants and the chances of data leakage between these tenants are very high. However, most of the service providers run a multi-tenant infrastructure. Proper investigations at the time of choosing the service provider must be done in order to avoid any such risk [97, 98].

3) In case a Cloud Service Provider uses a third party vendor to provide its cloud services, it should be ensured what service level agreements they have in between as well as what are the contingency plans in case of the breakdown of the third party system.

4) Proper SLAs defining the security requirements such as what level of encryption data should undergo, when it is sent over the internet and what are the penalties in case the service provider fails to do so.

Although data is stored outside the confines of the client organization in a public cloud, we cannot deny the possibility of an insider attack originating from service provider's end. Moving the data to a cloud computing environment expands the circle of insiders to the service provider's staff and sub-contractors [34]. An access control policy based on the inputs from the client and provider to prevent insider attacks has been proposed in [99]. Policy enforcement implemented at the nodes and the data-centres can prevent a system administrator from carrying out any malicious action. The three major steps to achieve this are: defining a policy, propagating the policy by means of a secure policy propagation module and enforcing it through a policy enforcement module.

## 6.2 Security issues in a private cloud

A private cloud model enables the customer to have total control over the network and provides the flexibility to the customer to implement any traditional network perimeter security practice. Although the security architecture is more reliable in a private cloud, yet there are issues/risks that need to be considered:

1) Virtualization techniques are quite popular in private clouds. In such a scenario, risks to the hypervisor should be carefully analyzed. There have been instances when a guest operating system has been able to run processes on other guest VMs or host. In a virtual environment it may happen that virtual machines are able to communicate with all the VMs including the ones who they are not supposed to. To ensure that they only communicate with the ones which they are supposed to, proper authentication and encryption techniques such as IPsec [IP level Security] etc. should be implemented [100].

2) The host operating system should be free from any sort of malware threat and monitored to avoid any such risk [101]. In addition, guest virtual machines should not be able to communicate with the host operating system directly. There should be dedicated physical interfaces for communicating with the host.

3) In a private cloud, users are facilitated with an option to be able to manage portions of the cloud, and access to the infrastructure is provided through a web-interface or an HTTP end point. There are two ways of implementing a web-interface, either by writing a whole application stack or by using a standard applicative stack, to develop the web interface using common languages such as Java, PHP, Python etc. As part of screening process, Eucalyptus web interface has been found to have a bug, allowing any user to perform internal port scanning or HTTP requests through the management node which he should not be allowed to do. In the nutshell, interfaces need to be properly developed and standard web application security techniques need to be deployed to protect the diverse HTTP requests being performed [102].

4) While we talk of standard internet security, we also need to have a security policy in place to safeguard the system from the attacks originating within the organization. This vital point is missed out on most of the occasions, stress being mostly upon the internet security. Proper security guidelines across the various departments should exist and control should be implemented as per the requirements [101].

Thus we see that although private clouds are considered safer in comparison to public clouds, still they have multiple issues which if unattended may lead to major security loopholes as discussed earlier.

The hybrid cloud model is a combination of both public and private cloud and hence the security issues discussed with respect to both are applicable in case of hybrid cloud. A trust model of cloud security in terms of social security has been discussed in [103]. Social insecurity has been classified as multiple stakeholder problem, open space security problem and mission critical data handling problem. All these issues have been considered while proposing a cloud trust model also known as "Security Aware Cloud". Two additional layers of trust: internal trust layer and contracted trust layer have been proposed to enhance security in a cloud computing environment.

## 7. ENSURING SECURITY AGAINST THE VARIOUS TYPES OF ATTACKS

In order to secure cloud against various security threats such as: *SQL injection, Cross Site Scripting (XSS), DoS and DDoS attacks, Google Hacking, and Forced Hacking*, different cloud service providers adopt different techniques. A few standard techniques to detect the above mentioned attacks include: avoiding the usage of dynamically generated SQL in the code, finding the meta-structures used in the code, validating all user entered parameters, and disallowing and removal of unwanted data and characters, etc. A comparative analysis of some of the currently existing security schemes has been done in Table 1.

A generic security framework needs to be worked out for an optimized cost performance ratio. The main criterion to be fulfilled by the generic security framework is to interface with any type of cloud environment, and to be able to handle and detect predefined as well as customized security policies.

A similar approach is being used by Symantec Message Labs Web Security cloud that blocks the security threats originating from internet and filters the data before they reach the network. Web security cloud's security architecture rests on two components:

- *Multi layer security:* In order to ensure data security and block possible malwares, it consists of multi-layer security and hence it has a strong security platform.
- *URL filtering:* It is being observed that the attacks are launched through various web pages and internet sites and hence filtering of the web-pages ensures that no such harmful or threat carrying web pages are accessible. Also, content from undesirable sites can be blocked.

With its adaptable technology, it provides security even in highly conflicting environments and ensures protection against new and converging malware threats.

The security model of Amazon Web Services, one of the biggest cloud service providers in the market makes use of multi-factor authentication technique, ensuring enhanced control over AWS account settings and the management of AWS services and resources for which the account is subscribed. In case the customer opts for Multi Factor Authentication (MFA), he has to provide a 6-digit code in addition to their username and password before access is granted to AWS account or services. This single use code can be received on mobile devices every time he tries to login into his/her AWS account. Such a technique is called multi-factor authentication, because two factors are checked before access is granted [64,104].

A Google hacking database identifies the various types of information such as: login passwords, pages containing logon portals, session usage information etc. Various software solutions such as Web Vulnerability Scanner can be used to detect the possibility of a Google hack. In order to prevent Google hack, users need to ensure that only those information that do not affect them should be shared with Google. This would prevent sharing of any sensitive information that may result in adverse conditions.

The symptoms of a *DoS or DDoS attack* are: system speed gets reduced and programs run very slowly, large number of connection requests from a large number of users, less number of available resources. Although, when launched in full strength, DDoS attacks are very harmful as they exhaust all the network resources, still a careful monitoring of the network can help in keeping these attacks in control [105]. An approach based on game theory against bandwidth consuming DoS and DDoS attacks has been proposed in [106]. The authors have modelled the interaction between the attacker and the user as a two player non-zero sum game in two attack scenarios a) single attacking node for DoS and b) multiple attacking nodes for DDoS attack. Based on these two scenarios, user is supposed to determine firewall settings to block unauthorized requests while allowing the authorized ones [106].

In case of *IP spoofing* an attacker tries to spoof authorized users creating an impression that the packets are coming from reliable sources. Thus the attacker takes control over the client's data or system showing himself/herself as the trusted party. Spoofing attacks can be checked by using encryption techniques and performing user authentication based on Key exchange. Techniques like *IPSec* do help in mitigating the risks of spoofing. By enabling encryption for sessions and performing filtering for incoming and outgoing packets, spoofing attacks can be reduced.

**Table 1. Comparative Analysis for Strengths and Limitations of Some of the Existing Security Schemes**

| Security Scheme | Suggested Approach | Strengths | Limitations |
|---|---|---|---|
| Data Storage security [48] | Uses homomorphic token with distributed verification of erasure-coded data towards ensuring data storage security and locating the server being attacked. | 1. Supports dynamic operations on data blocks such as: update, delete and append without data corruption and loss. 2. Efficient against data modification and server colluding attacks as well as against byzantine failures. | The security in case of dynamic data storage has been considered. However, the issues with fine-grained data error location remain to be addressed. |
| User identity safety in cloud computing | Uses active bundles scheme, whereby predicates are compared over encrypted data and multiparty computing. | Does not need trusted third party (TTP) for the verification or approval of user identity. Thus the user's identity is not disclosed. The TTP remains free and could be used for other purposes such as decryption. | Active bundle may not be executed at all at the host of the requested service. It would leave the system vulnerable. The identity remains a secret and the user is not granted permission to his requests. |
| Trust model for interoperability and security in cross cloud [81] | 1. Separate domains for providers and users, each with a special trust agent. 2. Different trust strategies for service providers and customers. 3. Time and transaction factors are taken into account for trust assignment. | 1. Helps the customers to avoid malicious suppliers. 2. Helps the providers to avoid cooperating/serving malicious users. | Security in a very large scale cross cloud environment is an active issue. This present scheme is able to handle only a limited number of security threats in a fairly small environment. |
| Virtualized defence and reputation based trust management | 1. Uses a hierarchy of DHT-based overlay networks, with specific tasks to be performed by each layer. 2. Lowest layer deals with reputation aggregation and probing colluders. The highest layer deals with various attacks. | Extensive use of virtualization for securing clouds. | The proposed model is in its early developmental stage and needs further simulations to verify the performance. |
| Secure virtualization [83] | 1. Idea of an Advanced Cloud Protection system (ACPS) to ensure the security of guest virtual machines and of distributed computing middleware is proposed. 2. Behaviour of cloud components can be monitored by logging and periodic checking of executable system files. | A virtualized network is prone to different types of security attacks that can be launched by a guest VM. An ACPS system monitors the guest VM without being noticed and hence any suspicious activity can be blocked and system's security system notified. | System performance gets marginally degraded and a small performance penalty is encountered. This acts as a limitation towards the acceptance of an ACPS system. |
| Safe, virtual network in cloud environment [81] | Cloud Providers have been suggested to obscure the internal structure of their services and placement policy in the cloud and also to focus on side-channel risks in order to reduce the chances of information leakage. | Ensures the identification of adversary or the attacking party and helping us find a far off place for an attacking party from its target and hence ensuring a more secure environment for the other VMs. | If the adversary gets to know the location of the other VMs, it may try to attack them. This may harm the other VMs in between. |

Every cloud service provider has installed various security measures depending on its cloud offering and the architecture. Their security model largely depends upon the customer section being served, type of cloud offering they provide and the deployment models they basically implement as discussed in [107].

One of the security measures implemented by SalesForce.com to avoid unauthorized access to its platform is sending a security code to the registered customer every-time the same account is accessed from same or different IP-address and the user needs to provide the security code at the time of logging in, in order to prove his/her identity [108].

It is equally important to secure the data-in-transit and security of transmitted data can be achieved through various encryption and decryption schemes. Steganography is another such technique which can be used to hide confidential information within a plain text, image, audio/video files or even IP datagrams in a TCP/IP network. A detailed analysis of various steganographic techniques and their application using different cover media has been done in [109].

Security issues in a virtualized environment wherein a malicious virtual machine tries to take control of the hypervisor and access the data belonging to other VMs have been observed and since traffic passing between VMs does not travel out into the rest of the data-centre network it cannot be seen by regular network based security platforms [110].

Hence, there is a need to ensure that security against the virtual threats should also be maintained by adopting the methodologies such as: checking the virtual machines connected to the host system and constantly monitoring their activity, securing the host computers to avoid tampering or file modification when the virtual machines are offline, preventing attacks directed towards taking control of the host system or other virtual machines on the network etc. Detecting unauthorized access in a cloud using tenant profiles can be performed as per the techniques mentioned in [111]. It has been assumed that the usage patterns in virtual machines in a cloud computing environment can be used to detect intrusions through abnormal usage. The recorded usage patterns over a period of time can be compared with the expected usage patterns which have already been provided by the tenants and deviations in case of unauthorized access can be detected.

A security model wherein a dedicated monitoring system tracking data coming in and out of a virtual machine/machines in a virtualized environment on a hypervisor can be presented as shown below:

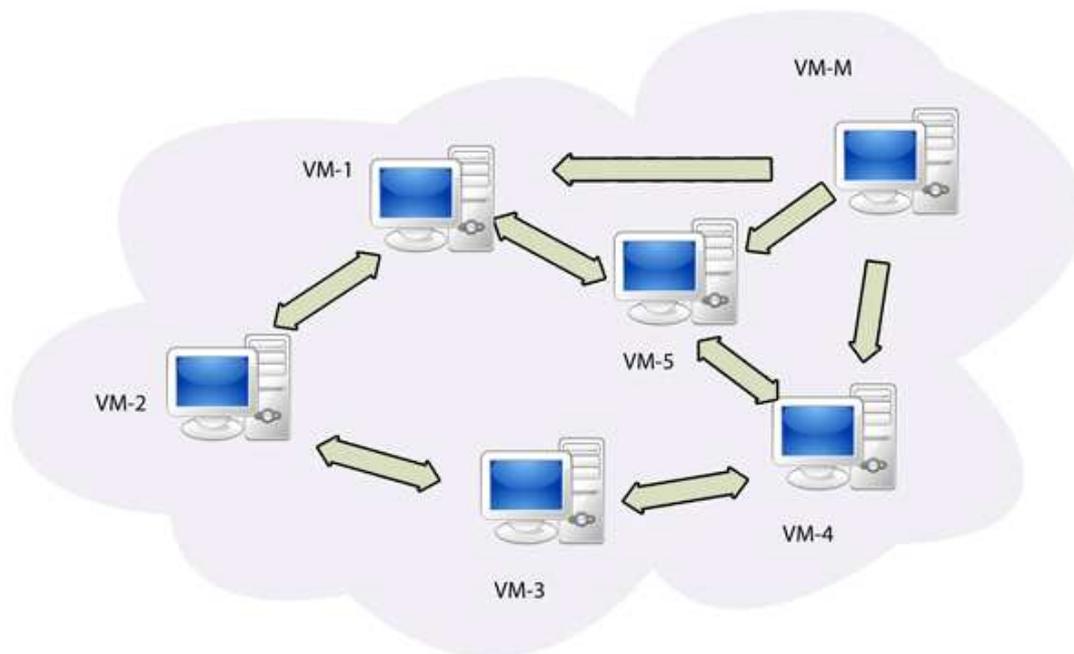

**Fig 2: Security Model in a Virtualized Environment**

As can be seen from the above shown security model [Fig. 2], a virtual machine monitor can be placed in a virtual environment which will keep track of all the traffic flowing in and out of a virtual machine network. And in case there is any suspicious activity observed, the corresponding virtual machine may be de-linked or blocked to maintain the security of the virtualized network.

The security breach of Twitter and Vaserv.com (via a zero-day vulnerability) and the data breach at Sony Corporation and Go-Grid [112], compromising 100 million customers' data [113] have made it quite clear that stringent security measures are needed to be taken in order to ensure security and proper data control in the cloud.

Thus we see that the security model adopted by a Cloud service provider should safeguard the cloud against all the possible threats and ensure that the data residing in the cloud does not get lost due to some unauthorized control over the network by some third party intruder.

## 8. CONCLUSION

The way cloud has been dominating the IT market, a major shift towards the cloud can be expected in the coming years. Already organizations have started moving into the cloud and a few of them includes: Schneider Electric implementing their CRM solutions on salesforce.com SaaS platform, Japanese automaker Toyota's pact with Microsoft to develop a new content delivery network for its automobiles on the latter's Azure cloud computing platform. More and more IT organizations will be moving into the cloud and with the emergence of NoSQL built around the technologies like Hadoop/HBase and Cassandra [114], collecting and using massive amount of data is no more considered as a headache. Questions such as how a cloud infrastructure is built will be superseded by how and in what way, to better utilize the cloud. Enhanced cross cloud connectivity and integration wherein different cloud deployment models will be integrated to provide a better infrastructure with feasible data migration options. Increasing tablet use and file based collaboration techniques will give way to cloud based service deployment models and an increased user-base in the cloud [115]. Technologies like Ruby on rails, HTML5 will continue to improve cloud experience in comparison to legacy options. Mobile cloud computing is expected to emerge as one of the biggest market for cloud service providers and cloud developers. Split processing techniques will come into picture and will be an enabling platform for mobile devices.

Although cloud computing has revolutionized the computing world, it is prone to a number of security threats varying from network level threats to application level threats. In order to keep the Cloud secure, these security threats need to be controlled. Moreover data residing in the cloud is also prone to a number of threats and various issues like: confidentiality and integrity of data should be considered while buying storage services from a cloud service provider. Auditing of the cloud at regular intervals needs to be done to safeguard the cloud against external threats. In addition to this, cloud service providers must ensure that all the SLA's are met and human errors on their part should be minimized, enabling smooth functioning. In this paper various security concerns related to the three basic services provided by a Cloud computing environment are considered and the solutions to prevent them have been discussed.

## 9. ACKNOWLEDGEMENTS
The authors are extremely thankful to Nabendu Chaki, Rituparna Chaki and Chandrakant Sakharwade for their invaluable suggestions towards the improvement of the paper.

## 10. REFERENCES

[1] L. Wang, Gregor Laszewski, Marcel Kunze, Jie Tao, "Cloud Computing: A Perspective Study", New Generation Computing- Advances of Distributed Information Processing, pp. 137-146, vol. 28, no. 2, 2008.
DOI: 10.1007/s00354-008-0081-5

[2] R. Maggiani, Communication Consultant, Solari Communication, "Cloud Computing is Changing How we Communicate", 2009 IEEE International Professional Conference, IPCC, pp. 1-4, Waikiki, HI, USA, July 19- 22, 2009. ISBN: 978-1-4244-4357-4.

[3] Harold C. Lin, Shivnath Babu, Jeffrey S. Chase, Sujay S. Parekh, "Automated Control in Cloud Computing: Opportunities and Challenges", Proc. of the 1st Workshop on Automated control for data centres and clouds, New York, NY, USA, pp. 13-18, 2009, ISBN: 978-1-60558-585-7.

[4] Peter Mell, Timothy Grance, "The NIST Definition of Cloud Computing", Jan, 2011.
http://docs.ismgcorp.com/files/external/Draft-SP-800-145_cloud-definition.pdf

[5] Meiko Jensen, Jorg Schwenk, Nils Gruschka, Luigi Lo Iacon, "On technical Security Issues in Cloud Computing", Proc. of IEEE International Conference on Cloud Computing (CLOUD-II, 2009), pp. 109-116, India, 2009.

[6] B.P. Rimal, Choi Eunmi, I. Lumb, "A Taxonomy and Survey of Cloud Computing Systems", Intl. Joint Conference on INC, IMS and IDC, 2009, pp. 44-51, Seoul, Aug, 2009.
DOI: 10.1109/NCM.2009.218

[7] Gaoyun Chen, Jun Lu and Jian Huang, Zexu Wu, "SaaAS - The Mobile Agent based Service for Cloud Computing in Internet Environment", Sixth International Conference on Natural Computation, ICNC 2010, pp. 2935-2939, IEEE, Yantai, Shandong, China, 2010. ISBN: 978-1-4244-5958-2.

[8] Sangeeta Sen, Rituparna Chaki, "Handling Write Lock Assignment in Cloud Computing Environment", Communications in Computer and Information Science, vol. 245, issue. 7, pp. 221-230, 2011.
DOI: 10.1007/978-3-642-27245-5_27

[9] Seny Kamara, Kristin Lauter, "Cryptographic cloud storage", Lecture Notes in Computer Science, Financial Cryptography and Data Security, pp. 136-149, vol. 6054, 2010.
DOI: 10.1007/978-3-642-14992-4_13

[10] S. Bhardwaj, L. Jain, and S. Jain, "Cloud computing: A study of infrastructure as a service (IAAS)", International Journal of engineering and information Technology, 2(1):60–63, 2010.

[11] R. L. Grossman, "The Case for Cloud Computing", IT Professional, vol. 11(2), pp. 23-27, Mar-April, 2009,

ISSN: 1520-9202, INSPEC Accession Number: 10518970, DOI: 10.1109/MITP.2009.40.

[12] Timothy Wood, Prashant Shenoy, Alexandre Gerber, K.K. Ramkrishnan, Jacobus Van der Merwe, "The Case for Enterprise-Ready Virtual Private Clouds", HotCloud'09 Proceedings of the 2009 conference on Hot topics in cloud computing, San Diego, CA, USA, 2009.
http://www.usenix.org/event/hotcloud09/tech/full_papers/wood.pdf

[13] Hoang T. Dinh, Chonho Lee, Dusit Niyato, Ping Wang, "A Survey of Mobile Cloud Computing: Architecture, Applications and Approaches", Wireless Communications and Mobile Computing, Wiley Journals, Oct 11, 2011.
DOI: 10.1002/wcm.1203

[14] Lizhe Wang, Jie Tao, Kunze M., Castellanos A.C., Kramer D., Karl W., "Scientific Cloud Computing: Early Definition and Experience", 10th IEEE Int. Conference on High Performance Computing and Communications, pp. 825-830, Dalian, China, Sep. 2008, ISBN: 978-0-7695-3352-0.

[15] Shuai Zhang, Shufen Zhang, Xuebin Chen, Xiuzhen Huo, "Cloud Computing Research and Development Trend", Intl. Conference on Future Networks, pp. 93-97, China, 2010.
DOI: 10.1109/ICFN.2010.58

[16] Aman Bakshi, Yogesh B. Dujodwala, "Securing cloud from DDoS Attacks using Intrusion Detection System in Virtual Machine", ICCSN '10 Proceeding of the 2010 Second International Conference on Communication Software and networks, pp. 260-264, 2010, IEEE Computer Society, USA, 2010. ISBN: 978-0-7695-3961-4.

[17] Youseff, L; Butrico, M; Da Silva, D., "Toward a Unified Ontology of Cloud Computing", Grid Computing Environments Workshop, pp. 1-10, Nov, 2008, Austin, Texas.
DOI: 10.1109/GCE.2008.4738443

[18] James Governor, "Web 2.0 Architectures: What Entrepreneurs and Information Architects Need to Know by James Governor", May 15, 2009; O'Reilly; ISBN-13: 978-0596514433.

[19] Amy Shuen, "Web 2.0: A Strategy Guide: Business thinking and strategies behind successful Web 2.0 implementations", O'Reilly Media; 1st edition; Apr 30, 2008; ISBN-13: 978-0596529963.

[20] Sam Murugesan, "Understanding Web 2.0", IEEE Computer Society, pp. 34-41. July-Aug, 2007.
http://91-592-722.wiki.uml.edu/file/view/understanding_web_20.pdf

[21] Antero Taivalsaari, "Mashware: The Future of Web Applications", Technical Report, Feb 2009.
http://labs.oracle.com/techrep/2009/smli_tr-2009-181.pdf
DOI: 10.1145/1878537.1878703

[22] David Chappel, "A Short Introduction to Cloud Platforms: An Enterprise Oriented View", David Chappel and Associates, August, 2008. [Sponsored by Microsoft Corporation]
http://www.davidchappell.com/CloudPlatforms--Chappell.pdf

[23] Qi Zhang, Lu Cheng, Raouf Boutaba, "Cloud Computing: State of the art and research challenges", Journal of Internet Services and Applications, pp. 7-18, vol. 1, issue. 1, Feb, 2010.
DOI: 10.1007/s13174-010-0007-6

[24] R. Gellman, "Privacy in the clouds: Risks to privacy and confidentiality from cloud computing", The World Privacy Forum, 2009.
http://www.worldprivacyforum.org/pdf/WPF_Cloud_Privacy_Report.pdf.

[25] Tim Mather, Subra Kumaraswamy, Shahed Latif, "Cloud Security and Privacy: An Enterprise Edition on Risks and Compliance (Theory in Practice)", O'Reilly Media, Sep. 2009; ISBN: 978-0596802769.
http://oreilly.com/catalog/9780596802776.

[26] Lori M. Kaufman, "Data security in the world of cloud computing", IEEE Security and Privacy Journal, vol. 7, issue. 4, pp. 61-64, July- Aug 2009, ISSN: 1540-7993, INSPEC Accession Number: 10805344, DOI: 10.1109/MSP.2009.87.

[27] Md Tanzim Khorshed, A. B. M. Shawkat Ali, Saleh A. Wasimi, "Trust Issues that create threats for Cyber attacks in Cloud Computing", IEEE 17th International Conference on Parallel and Distributed Systems, pp. 900-905, 2011.

[28] S. Pearson, "Taking account of privacy when designing cloud computing services", CLOUD '09 Proc. of ICSE Workshop on Software Engineering Challenges of Cloud Computing, pp. 44-52, IEEE Computer Society Washington, DC, USA, May 2009. ISBN: 978-1-4244-3713-9.

[29] George V. Hulme, "NIST formalizes cloud computing definition, issues security and privacy guidance", Feb. 3, 2011 [A common platform enabling security executives to share best security practices and strategic insights].
http://www.csoonline.com/article/661620/nist-formalizes-cloud-computing-definition-issues-security-and-privacy-guidance.

[30] Julisch, K., & Hall, M., "Security and control in the cloud", Information Security Journal: A Global Perspective, vol. 19, no. 6, pp. 299-309, 2010.

[31] Chi-Chun Lo, Chun-Chieh Huang, Joy Ku, "A Cooperative Intrusion Detection System Framework for Cloud Computing Networks", ICPPW '10 Proceedings of the 2010 39th International Conference on Parallel Processing Workshops, IEEE Computer Society, pp. 280-284, Washington DC, USA, 2010. ISBN: 978-0-7695-4157-0.

[32] Hamid R. Motahari-Nezhad, Claudio Bartolini, Sven Graupner, Sharad Singhal, Susan Spence, "IT Support Conversation Manager: A Conversation-Centered Approach and Tool for Managing Best Practice IT Processes", Proceedings of the 2010 14th IEEE International Enterprise Distributed Object Computing Conference, pp. 247-256, October 25-29, 2010, ISBN: 978-1-4244-7966-5.

[33] L.J. Zhang and Qun Zhou, "CCOA: Cloud Computing Open Architecture", ICWS 2009: IEEE International Conference on Web Services, pp. 607-616. July 2009. DOI: 10.1109/ICWS.2009.144.

[34] Wayne Jansen, Timothy Grance, "NIST Guidelines on Security and Privacy in Public Cloud Computing", Draft Special Publication 800-144, 2011. http://csrc.nist.gov/publications/drafts/800-144/Draft-SP-800-144_cloud-computing.pdf.

[35] Jon Marler, "Securing the Cloud: Addressing Cloud Computing Security Concerns with Private Cloud", Rackspace Knowledge Centre, March 27, 2011, Article Id: 1638. http://www.rackspace.com/knowledge_center/private-cloud/securing-the-cloud-addressing-cloud-computing-security-concerns-with-private-cloud

[36] Frederik De Keukelaere, Sumeer Bhola, Michael Steiner, Suresh Chari, Sachiko Yoshihama, "Smash: secure component model for cross-domain mashups on unmodified browsers", Proc. of the 17th International Conference on World Wide Web, ACM, NY, USA, 2008, ISBN: 978-1-60558-085-2, DOI: 10.1145/1367497.1367570.

[37] Michael Armbrust, Armando Fox, Rean Griffith, Anthony D. Joseph, Randy Katz, Andy Konwinski, Gunho Lee, David Petterson, Ariel Rabkin, Ion Stoica, Matei Zaharica, "A View of Cloud Computing", Communications of the ACM, vol. 53, issue. 4, April 2010, USA. DOI: 10.1145/1721654.1721672

[38] Neal Leavitt, "Is Cloud Computing Really Ready for Prime Time?" Computer, vol. 42, issue. 1, pp. 15-20, IEEE Computer Society, CA, USA, January 2009. ISSN: 0018-9162.

[39] Robert Minnear, "Latency: The Achilles Heel of Cloud Computing", March 9, 2011, Cloud Expo: Article, Cloud Computing Journal. http://cloudcomputing.sys-con.com/node/1745523.

[40] Daniele Catteddu, Giles Hogben, "Cloud Computing: Benefits, Risks and Recommendations for Information Security", European Network and Information Security Agency (ENISA), Nov, 2009. http://www.enisa.europa.eu/act/application-security/test/act/rm/files/deliverables/cloud-computing-risk-assessment

[41] Marios D. Dikaiakos, Dimitrios Katsaros, Pankaj Mehra, George Pallis, Athena Vakali, "Cloud Computing: Distributed Internet Computing for IT and Scientific Research", IEEE Internet Computing Journal, vol. 13, issue. 5, pp. 10-13, September 2009. DOI: 10.1109/MIC.2009.103.

[42] Michael Kretzschmar, S Hanigk, "Security management interoperability challenges for collaborative clouds", Systems and Virtualization Management (SVM), 2010, Proceedings of the 4th International DMTF Academic Alliance Workshop on Systems and Virtualization Management: Standards and the Cloud, pp. 43-49, October 25-29, 2010. ISBN: 978-1-4244-9181-0, DOI: 10.1109/SVM.2010.5674744.

[43] B. R. Kandukuri, R. V. Paturi and A. Rakshit, "Cloud Security Issues", 2009 IEEE International Conference on Services Computing, Bangalore, India, September 21-25, 2009. In Proceedings of IEEE SCC'2009. pp. 517-520, 2009. ISBN: 978-0-7695-3811-2.

[44] Jessica T., "Connecting Data Centres over Public Networks", IPEXPO.ONLINE, April 20, 2011. http://online.ipexpo.co.uk/2011/04/20/connecting-data-centres-over-public-networks/

[45] Cong Wang, Kui Ren, Wenjing Lou, Jin Li, "Towards Publicly Auditable Secure Cloud Storage Services", IEEE Networks, pp. 19-24, vol. 24, issue. 4, July, 2010. DOI: 10.1109/MNET.2010.5510914

[46] H. Liang, D. Huang, L. X. Cai, X. Shen and D. Peng, "Resource allocation for security services in mobile cloud computing", in Proc. IEEE INFOCOM'11, Machine-to-Machine Communications and Networking (M2MCN), pp. 191-195, April 10-15, 2011, Shanghai, China.

[47] Martin Mulazzani, Sebastian Schrittwieser, Manuel Leithner, Markus Huber, Edgar Weippl, "Dark Clouds on the Horizon: Using Cloud Storage as Attack Vector and Online Slack Space", Proceedings of the 20th USENIX conference on Security, Berkley, USA, 2011.

[48] Cong Wang, Qian Wang, Kui Ren, and Wenjing Lou, "Ensuring Data Storage Security in Cloud Computing", 17th International workshop on Quality of Service,2009, IWQoS, Charleston, SC, USA, pp.1-9, July 13-15, 2009, ISBN: 978-1-4244-3875-4.

[49] W. Li, L. Ping, X. Pan, "Use trust management module to achieve effective security mechanisms in cloud environment", 2010 International Conference on Electronics and Information Engineering (ICEIE), Volume: 1, pp. V1-14 - V1-19, 2010. DOI: 10.1109/ICEIE.2010.5559829.

[50] R. A. Vasudevan, A. Abraham, S.Sanyal, D.P. Agarwal, "Jigsaw-based secure data transfer over computer networks", Int. Conference on Information Technology: Coding and Computing, pp. 2-6, vol.1, April, 2004.

[51] R. A. Vasudevan, S. Sanyal, "A Novel Multipath Approach to Security in Mobile Ad Hoc Networks (MANETs)", Int. Conference on Computers and Devices for Communication, CODEC'04, Kolkata, India.

[52] Jeff Sedayao, Steven Su, Xiaohao Ma, Minghao Jiang and Kai Miao, "A Simple Technique for Securing Data at Rest", Lecture Notes in Computer Science, pp. 553-558, 2009. DOI: 10.1007/978-3-642-10665-1_51

[53] Yogesh L. Simmhan, Beth Plale, Dennis Gannon, "A Survey of Data Provenance Techniques", ACM SIGMOD, vol. 34, issue. 3, Sep, 2005, NY, USA. DOI: 10.1145/1084805.1084812

[54] P. R. Gallagher, "Guide to Understanding Data Remanence in Automated Information Systems", The Rainbow Books, ch3 and ch.4, 1991.

[55] Larry Dignan (Editor in Chief- ZDNet), "Epsilon Data Breach: What's the value of an email address", IT Security Blogs, Tech Republic, April 5, 2011. http://www.techrepublic.com/blog/security/epsilon-data-breach-whats-the-value-of-an-email-address/5307

[56] Farzad Sabahi, "Secure Virtualization for Cloud Environment Using Hypervisor-based Technology", Int. Journal of Machine Learning and Computing, pp. 39-45, vol. 2, no. 1, February, 2012.

[57] David Goldman, "Why Amazon's Cloud Titanic Went Down", CNNMoney, April, 2011. http://money.cnn.com/2011/04/22/technology/amazon_ec2_cloud_outage/index.htm

[58] Rory Smith (SOC Analyst), "The Use of Legitimate Channels to distribute malicious software to Users", Security Samurai, Aug. 2, 2011. http://www.thesecuritysamurai.com/2011/08/02/the-use-of-legitimate-channels-to-distribute-malicious-software-to-users-by-rory-smith-soc-analyst/

[59] Thomas Ristenpart, Eran Tromer, Hovav Shacham, Stefan Savage, "Hey, you get off my cloud: Exploring information leakage in third party compute clouds", CCS'09, Proceedings of the 16th ACM conference. On Computer and Communications Security, pp. 199-212, ACM New York, NY, USA, 2009. ISBN: 978-1-60558-894-0.

[60] Michael Krigsman, "MediaMax/The Linkup: When the Cloud fails", IT Project Failures, News and Blogs, ZDNet, August, 2008. http://www.zdnet.com/blog/projectfailures/mediamax-the-linkup-when-the-cloud-fails/999

[61] K. Hwang, S Kulkarni and Y. Hu, "Cloud security with virtualized defence and Reputation-based Trust management", Proceedings of 2009 Eighth IEEE International Conference on Dependable, Autonomic and Secure Computing (security in cloud computing), pp. 621-628, Chengdu, China, December, 2009. ISBN: 978-0-7695-3929-4.

[62] Ryan K.L.Ko, Bu Sung Lee and Siani Pearson, "Towards Achieving Accountability, Auditability and Trust in Cloud Computing", Communications in Computer and Information Science, Vol. 193(4), pp. 432-444, 2011.
DOI: 10.1007/978-3-642-22726-4_45

[63] S. Subashini, V. Kavitha, "A survey on security issues in service delivery models of cloud computing", Journal of Network and Computer Applications, Vol. 34(1), pp 1–11, Academic Press Ltd., UK, 2011, ISSN: 1084-8045.

[64] "Amazon Web Services: Overview of Security Processes", Whitepaper, May, 2011. http://d36cz9buwru1tt.cloudfront.net/pdf/AWS_Security_Whitepaper.pdf

[65] Pradnyesh Rane, "Securing SaaS Applications: A Cloud Security Perspective for Application Providers", Information Security Management Handbook, Vol. 5, 2010. http://www.infosectoday.com/Articles/Securing_SaaS_Applications.htm

[66] Ruixuan Li, Li Nie, Xiaopu Ma, Meng Dong, Wei Wang, "SMEF: An Entropy Based Security Framework for Cloud-Oriented Service Mashup", Int. Conf on Trust, Security and Privacy in Computing and Communications (TrustCom), pp. 304-311, Nov, 2011.
DOI: 10.1109/TrustCom.2011.41

[67] Adam A Noureddine, Meledath Damodaran, "Security in Web 2.0 Application Development", iiWAS '08, Proc. of the 10th International Conference on Information Integration and Web-based Applications & Services, pp. 681-685, 2008, ISBN: 978-1-60558-349-5, DOI: 10.1145/1497308.1497443.

[68] Justin Clarke; SQL Injection Attacks and Defense; Syngress 2009; ISBN-13: 978-159749424.

[69] A. Liu, Y. Yuan, A Stavrou, "SQLProb: A Proxy-based Architecture towards Preventing SQL Injection Attacks", SAC March 8-12, 2009, Honolulu, Hawaii, U.S.A.

[70] P. Vogt, F. Nentwich, N. Jovanovic, E. Kirda, C. Kruegel, and G. Vigna, "Cross-Site Scripting Prevention with Dynamic Data Tainting and Static Analysis", Proceedings of the Network and Distributed System Security Symposium (NDSS'07), February, 2007.

[71] D. Gollmann, "Securing Web Applications", Information Security Technical Report, vol. 13, issue. 1, 2008, Elsevier Advanced Technology Publications Oxford, UK, DOI: 10.1016/j.istr.2008.02.002.

[72] Ter Louw, M; Venkatakrishnan, V. N.; "BluePrint: Robust Prevention of Cross-Site scripting attacks for existing browsers", 30th IEEE Symposium on Security and Privacy, pp. 331-346, May, 2009.
DOI: 10.1109/SP.2009.33

[73] Jonathan Katz, "Efficient Cryptographic Protocols Preventing Man in the Middle Attacks", Doctoral Dissertation submitted at Columbia University, 2002, ISBN: 0-493-50927-5. http://www.cs.ucla.edu/~rafail/STUDENTS/katz-thesis.pdf /

[74] Eric Ogren, "Whitelists SaaS modify traditional security, tackle flaws", Sep. 17, 2009. [Eric Ogren is the founder and principal security analyst at Ogren Group] http://searchsecurity.techtarget.com/news/column/0,294698,sid14_gci1368647,00.html/

[75] Gurdev Singh, Amit Sharma, Manpreet Singh Lehal, "Security Apprehensions in Different Regions of Cloud Captious Grounds", International Journal of Network Security & Its Applications (IJNSA), Vol.3, No.4, July 2011.

[76] Char Sample, Senior Scientist, BBN Technologies, Diana Kelley, Partner, Security Curve, "Cloud computing security: Routing and DNS security threats". http://searchsecurity.techtarget.com/tip/0,289483,sid14_gci1359155_mem1, 00.html/

[77] Zouheir Trabelsi, Hamza Rahmani, Kamel Kaouech, Mounir Frikha, "Malicious Sniffing System Detection Platform", Proceedings of the 2004 International Symposium on Applications and the Internet (SAINT'04), pp. 201-207, 2004, ISBN: 0-7695-2068-5.

[78] Josh Karlin, Stephanie Forrest, Jennifer Rexford, "Autonomous Security for Autonomous Systems", Proc. of Complex Computer and Communication Networks; vol. 52, issue. 15, pp. 2908- 2923, Oct. 2008, Elsevier North-Holland, Inc. New York, NY, USA.

[79] Scalable Security Solutions, Check Point Open Performance Architecture, Quad-Core Intel Xeon Processors, "Delivering Application-Level Security at Data Centre Performance Levels", Intel Corporation, Whitepaper, 2008. http://download.intel.com/netcomms/technologies/security/320923.pdf

[80] Shengmei Luo, Zhaoji Lin, Xiaohua Chen, Zhuolin Yang, Jianyong Chen, "Virtualization security for cloud computing services", Int. Conf on Cloud and Service Computing, pp. 174-179, Dec, 2011. DOI: 10.1109/CSC.2011.6138516

[81] Shantanu Pal, Sunirmal Khatua, Nabendu Chaki, Sugata Sanyal, "A New Trusted and Collaborative Agent Based Approach for Ensuring Cloud Security", Annals of Faculty Engineering Hunedoara International Journal of Engineering (Archived copy), scheduled for publication in vol. 10, issue 1, January 2012. ISSN: 1584-2665.

[82] Jenni Susan Reuben, "A Survey on Virtual Machine Security", Seminar of Network Security, Helsinki University of Technology, 2007. http://www.tml.tkk.fi/Publications/C/25/papers/Reuben_final.pdf?q=attacks-on-virtual-machine-emulators

[83] Flavio Lombardi, Roberto Di Pietro, "Secure Virtualization for Cloud Computing", Journal of Network and Computer Applications, vol. 34, issue 4, pp. 1113- 1122, July 2011, Academic Press Ltd. London, UK.

[84] Hanqian Wu, Yi Ding, Winer, C., Li Yao, "Network Security for Virtual Machines in Cloud Computing", 5[th] Int'l Conference on Computer Sciences and Convergence Information Technology, pp. 18-21, Seoul, Nov. 30-Dec. 2, 2010. ISBN: 978-1-4244-8567-3.

[85] K. Vieira, A. Schulter, C. B. Westphall, and C. M. Westphall, "Intrusion detection techniques for Grid and Cloud Computing Environment", IT Professional, IEEE Computer Society, vol. 12, issue 4, pp. 38-43, 2010. DOI: 10.1109/MITP.2009.89.

[86] Ian Rathie, "An Approach to Application Security", SANS Security Essentials White Paper, SANS Institute. http://www.sans.org/reading_room/whitepapers/application/approach-application-security_16

[87] Ruiping Lua and Kin Choong Yow, "Mitigating DDoS Attacks with Transparent and Intelligent Fast-Flux Swarm Network", IEEE Network, vol. 25, no. 4, pp. 28-33, July-August, 2011.

[88] Nathan Mcfeters, "Recent CNN Distributed Denial of Service (DDoS) Attack Explained", ZDNet, April, 2008. http://www.zdnet.com/blog/security/recent-cnn-distributed-denial-of-service-ddos-attack-explained/1054

[89] Claudio Mazzariello, Roberto Bifulco and Roberto Canonico, "Integrating a Network IDS into an Open Source Cloud Computing Environment", Sixth International Conference on Information Assurance and Security, USA, pp. 265-270, Aug. 23-25, 2010. DOI: 10.1109/ISIAS.2010.5604069.

[90] D. Nurmi, R. Wolski, C. Grzegorczyk, G. Obertelli, S. Soman, L. Youseff, and D. Zagorodnov, "The Eucalyptus open-source cloud-computing system", in Proceedings of the 9th IEEE/ACM International Symposium on Cluster Computing and the Grid (CCGRID '09), pp. 124–131, 2009.

[91] John E. Dunn, "Spammers break Hotmail's CAPTCHA yet again", Tech-world, Feb. 16, 2009. http://news.techworld.com/security/110908/spammers-break-hotmails-captcha-yet-again/

[92] Albert B Jeng, Chien Chen Tseng, Der-Feng Tseng, Jiunn-Chin Wang, "A Study of CAPTCHA and its Application to User Authentication", Proc. Of 2[nd] Intl. Conference on Computational Collective Intelligence: Technologies and Applications, 2010. ISBN: 3-642-16731-4 978-3-642-16731-7

[93] Cui, JingSong; Wang, LiJing; Mei, JingTing; Zhang, Da; Wang, Xia; Peng, Yang; Zhang, WuZhou; "CAPTCHA design based on moving object recognition problem", Intl. Conference on Information Sciences and Interaction Sciences, pp. 158-162, June, 2010, China. DOI: 10.1109/ICICIS.2010.5534730

[94] V. Kumar, M. Singh, A. Abraham, S. Sanyal, "CompChall: addressing password guessing attacks", Int. Conference on Information Technology: Coding and Computing, pp. 739-744, vol. 1, April, 2005.

[95] Kellep Charles, "Google's Gmail Hacked by China Again", SecurityOrb, The Information Security knowledge-Base Website, June 2, 2011. http://securityorb.com/2011/06/googles-gmail-hacked-by-china-again/

[96] Ningning Zhu, Tzi-cker Chiueh, "Portable and Efficient Continuous Data Protection for Network File Servers", Intl. Conference on Dependable Systems and Networks, pp. 687-697, Edinburg, June, 2007. DOI: 10.1109/DSN.2007.74

[97] A. Verma and S. Kaushal, "Cloud Computing Security Issues and Challenges: A Survey", Proceedings of Advances in Computing and Communications, Vol. 193, pp. 445-454, 2011. DOI: 10.1007/978-3-642-22726-4_46

[98] P. Sharma, S. K. Sood, and S. Kaur, "Security Issues in Cloud Computing", Proceedings of High Performance Architecture and Grid Computing, Vol. 169, pp. 36-45, 2011. DOI: 10.1007/978-3-642-22577-2_5

[99] Sudharsan Sundararajan, Hari Narayanan, Vipin Pavithran, Kaladhar Vorungati, Krishnashree Achuthan, "Preventing Insider attacks in the Cloud", Communications in Computer and Information Science, vol. 190, issue. 5, pp. 488-500, 2011. DOI: 10.1007/978-3-642-22709-7_48

[100] Thomas W. Shinder, "Security Issues in Cloud Deployment models", TechNet Articles, Wiki, Microsoft, Aug, 2011. http://social.technet.microsoft.com/wiki/contents/articles/security-issues-in-cloud-deployment-models.aspx

[101] E. Mathisen, "Security Challenges and Solutions in Cloud Computing", Proceedings of the 5th IEEE International Conference on Digital Ecosystems and Technologies (DEST), pp. 208-212, June, 2011, ISBN: 978-1-4577-0871-8, DOI: 10.1109/DEST.2011.5936627.

[102] Alessandro Perilli, Claudio Criscione, "Securing the Private Cloud", Article on Secure Networks, Virtualization.info. http://virtualization.info/en/security/privatecloud.pdf

[103] Sato, H; Kanai, A; Tanimoto, S; "A Cloud Trust Model in a Security Aware Cloud", Intl. Symposium on Applications and the Internet (SAINT), pp. 121-124, July, 2010, Seoul.

[104] Ayu Tiwari, Sudip Sanyal, Ajith Abraham, Svein Johan Knapskog, Sugata Sanyal, "A Multi-Factor Security Protocol for Wireless Payment – Secure Web Authentication Using Mobile Devices", IADIS, International Conference Applied Computing, pp. 160-167, 2007.

[105] Tao Peng, Christopher Leckie, Kotagiri RamMohanRao, "Survey of Network Based Defense Mechanisms Countering the DoS and DDoS Problems", ACM Computing Surveys, vol. 39, no. 1, April, 2007. DOI: 10.1145/1216370.1216373

[106] Qishi Wu, Sajjan Shiva, Sankardas Roy, Charles Ellis, Vivek Datla, "On Modelling and Simulation of game-theory based defense mechanisms against DoS and DDoS attacks", Proceedings of 2010 Spring Simulation Multiconference, NY, USA, 0032010. DOI: 10.1145/1878537.1878703

[107] Amitav Chakravartty, Serena Software, "Serena Service Manager Security in the Cloud". http://www.serena.com/docs/repository/products/service-manager/Serena-Service-Manager-Security-in-the-Cloud.pdf

[108] Security and Privacy policies of sales-force.com, "Secure, Private and Trustworthy: Enterprise Cloud Computing with Force.com". http://www.salesforce.com/assets/pdf/misc/WP_Forcedotcom-Security.pdf http://trust.salesforce.com/trust/security/best_practices/ http://trust.salesforce.com/trust/privacy/tools/

[109] Soumyendu Das, Subhendu Das, Bijoy Bandopadhyay, Sugata Sanyal, "Steganography and Staganalysis: Different Approaches", Int. Journal of Computers, Information Technology and Engineering (IJCITAE), vol.2, no.1, June, 2008.

[110] Richard Chow, Philippe Golle, Markus Jakobsson, Elaine Shi, Jessicca Staddon, Ryusuke Masuoka, Jesus Molina, "Controlling Data in the Cloud: Outsourcing Computation without Outsourcing Control", Proc. of the ACM Workshop on Cloud Computing Security, pp. 85-90, USA, November, 2009. ISBN: 978-1-60558-784-4.

[111] Jason Nikolai, "Detecting Unauthorized Usage in a Cloud using Tenant Profiles". http://www.homepages.dsu.edu/malladis/teach/717/Papers/nikolai.pdf

[112] Craig Balding, "GoGrid Security Breach", cloudsecurity.org, March 30, 2011. http://cloudsecurity.org/blog/2011/03/30/gogrid-security-breach.html

[113] Czaroma Roman, "Sony Data Breach Highlights Importance of Cloud Security", Cloud Times, May 9, 2011. http://cloudtimes.org/sony-data-breach-highlights-importance-of-cloud-security/

[114] Hiroshi Wada, Alan Fekete, Liang Zhao, Kevin Lee, Anna Liu, "Data Consistency Properties and the Trade-offs in Commercial Cloud Storages : The Consumers' Perspective", Proc. of the 5th Biennial Conference on Innovative Data Systems Research (CIDR '2011), Asilomar, CA, January 2011.

[115] J. Weinman, "The Future of Cloud Computing", IEEE Technology Time Machine Symposium on Technologies Beyond 2020 (TTM), pp. 1-2, June, 2011. DOI: 10.1109/TTM.2011.6005157